# Tag-Pag: A Dedicated Tool for Systematic Web Page Annotations

Anton Pogrebnjak [1], Julian Schelb [1], Andreas Spitz [1], Celina Kacperski [2], and Roberto Ulloa [2]

[1]*Department of Computer Science, University of Konstanz, Germany*
[2]*Cluster of Excellence "The Politics of Inequalities", University of Konstanz, Germany*

26.08.2024

## Summary

Tag-Pag is an application designed to simplify the categorization of web pages, a task increasingly common for researchers who scrape web pages to analyze individuals' browsing patterns or train machine learning classifiers. Unlike existing tools that focus on annotating sections of text, Tag-Pag systematizes page-level annotations, allowing users to determine whether an entire document relates to one or multiple predefined topics.

Tag-Pag offers an intuitive interface to configure the input web pages and annotation labels. It integrates libraries to extract content from the HTML and URL indicators to aid the annotation process. It provides direct access to scraped live versions of the web page. Our tool is designed to expedite the annotation process with features like quick navigation, label assignment, and export functionality, making it a versatile and efficient tool for various research applications. Tag-Pag is available at https://github.com/Pantonius/TagPag

## Statement of need

The annotation of web data is increasingly common across multiple disciplines, serving purposes such as analyzing online behavioral patterns (Guess, 2021; Stier et al., 2020; Ulloa & Kacperski, 2023; Wojcieszak et al., 2024), auditing the performance of online platforms (Kacperski et al., 2024; Makhortykh et al., 2020), or training and evaluating machine learning classifiers (Schelb et al., 2024). As the need for processing web data grows, researchers have turned to systematic and efficient methodologies that span the entire process—from data collection to categorization—to ensure robust results.

Online behavioral researchers have recently investigated limitations of web scraping, such as reliance on external environments that differ from individuals' computers (Ulloa, Mangold, et al., 2024) and changes due to time delays in scraping (Dahlke et al., 2023; Ulloa, Mangold, et al., 2024). To improve reliability and validity, researchers attempt to scrape data from web pages as close to the visit time of participants as possible and uniformly distribute the delay between the visit and the web page collection. Such limitations have, more recently, been addressed by developing new web tools that collect content directly from an individual's browser (Adam et al., 2024; GESIS Panel Team, 2025). Other researchers have gone to the effort of standardizing web data collections for algorithm auditing to avoid, for example, noise stemming from search engine personalization (Ulloa, Makhortykh, et al., 2024).

These academic efforts illustrate the importance placed on collecting high-quality data for annotation purposes; so far, however, there has been a lack of tools to facilitate the manual annotation process. This has led researchers to instead use inefficient methods such as relying on the URL, rarely accessing the systematically scraped content, or manually visiting (and revisiting) the related web page at different times. As a result, promising lines of inquiry — especially those requiring large-scale and consistent annotations — might be left unexplored.

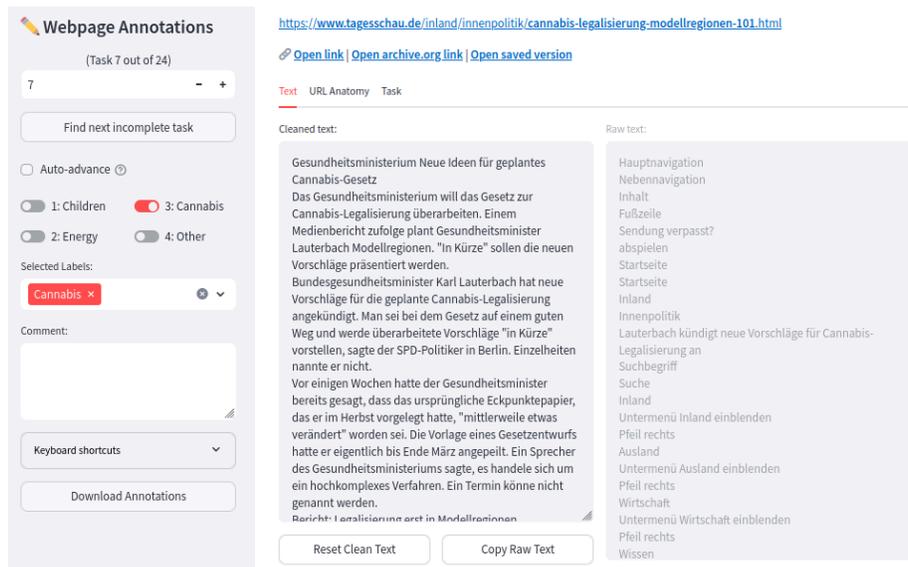

**Figure 1: Annotation interface for web pages classification tasks.** The interface of Tag-Pag is divided into a left sidebar and a main panel. The sidebar provides task navigation, annotation selection labels, and additional tools such as key shortcut references and annotation downloads. The main panel displays web page content in multiple views, including a cleaned text version, raw text, and URL decomposition. Annotators can label data using predefined categories and add comments. Tag-Pag automatically highlights relevant sections of the URL (at the top).

Existing tools often focus on annotating specific sections within a text (Huang, 2016; Meister, 2023; Rampin & Rampin, 2021), where the user selects a portion of text and assigns a label or establishes connections between parts of speech (Strippel et al., 2022). These tools fall short when the goal is to annotate entire pages to determine, for example, if the content corresponds to very specific topics (Schelb et al., 2024), misinformation (Urman et al., 2022), or, more broadly, political content (Guess, 2021; Stier et al., 2020), news articles (Ulloa & Kacperski, 2023) or pages that restrict access such as logins (Dahlke et al., 2023). Tag-Pag[1] addresses this gap by allowing broad-level annotations of entire web pages.

Tag-Pag uses libraries to extract two versions of the content from the HTML (see Figure 1): (1) cleaned text (Barbaresi, 2021), with removed boilerplate such as menus and advertisements, and (2) raw text (Artem Golubin, 2023), with only removed HTML elements. The tool also parses the URLs themselves, which often contain relevant information about the page's content, adding another layer of contextual data for annotations. For a comprehensive overview, users can open the scraped HTML, the live web page, or the latest version stored in the Wayback Machine. For researchers creating or refining training datasets to improve machine learning models, Tag-Pag allows easy text editing to retain only the relevant parts for classifiers, e.g., manually removing boilerplate to further filter the cleaned text.

Additionally, Tag-Pag includes functionality designed to speed up the annotation process: key bindings for interface actions for rapid label assignment, automatic transition between pages for single-label annotations are supported, and a feature to locate unannotated pages is included. Comments and annotations can be exported to CSV, ensuring compatibility with further steps of the analysis pipeline.

The tool also supports multiple annotators, with the functionality to hide one another's annotations and randomize the tasks' order to avoid priming effects (Mathur et al., 2017; Shen et al., 2019).

By integrating these features, Tag-Pag offers a systematic, efficient, and user-friendly approach to web page annotation, addressing the needs of researchers across various disciplines.

## Acknowledgments

This research was funded by the Deutsche Forschungsgemeinschaft (DFG – German Research Foundation) under Germany's Excellence Strategy – EXC 2035/1 – 390681379. The funders had no role in study design, data collection and analysis, decision to publish, or preparation of the manuscript.

---

[1] Tag-Pag: https://github.com/Pantonius/TagPag